\documentclass[aps,12pt]{revtex4} 
\usepackage{graphicx} 
\draft 

\begin{document}
\title{Overdamped quantum phase diffusion and charging effects in
  Josephson junctions}

\author{J.~Ankerhold}
\email{ankerhold@physik.uni-freiburg.de}
\affiliation{Physikalisches Institut, Albert-Ludwigs-Universit{\"a}t
Freiburg,
 Hermann-Herder-Stra{\ss}e 3, D-79104 Freiburg, Germany}

\date{February 20, 2004}


\begin{abstract}
Exploiting the recently derived quantum Smoluchowski equation the
classical Ivanchenko Zil'berman theory for overdamped diffusive phase
motion of  low capacitance Josephson junctions is extended to the low
temperature quantum domain where charging
effects appear. This formulation allows to derive
explicit results for the current-voltage characteristics over a broad
range of parameters that reduce to known findings in certain limits.
In particular, the transparent analytical approach comprises Coulomb
blockade physics, coherent Cooper
pair transfer, and the precursors of macroscopic quantum tunneling and
needs to be supplemented by more sophisticated methods only at very
low temperatures.
\end{abstract}
\maketitle

\section{Introduction}

In the last decades the physics of  Josephson junctions (JJ) has revealed
an extraordinary wealth of phenomena studied  theoretically and
experimentally as well \cite{barone,schon,nato}. The underlying system,
two superconducting domains separated by a tunnel barrier, can be found
in a variety of realizations, recently e.g.\ in superconducting atomic
contacts \cite{atomic} and solid-state quantum bits \cite{qubit}.
 Basically two parameters determine the dynamics, namely, the
Josephson coupling energy $E_{\rm J}$ and the charging energy $E_{\rm
c}=2 e^2/C$ of a junction with capacitance $C$. It turns out that the
competition between these two scales is crucially influenced by the
electromagnetic environment surrounding the junction, i.e.\ its impedance
which in the simplest case is given by an ohmic resistor with
resistance $R$.

For the classical JJ dynamics charging effects are negligible and the
charge transfer corresponds to a phase-coherent Cooper pair current
\begin{equation}
I_{\rm s}=I_{\rm c}\, \sin(\phi)\ \ \ \ ,\ \ \ \ \dot{\phi}(t)=\frac{2
e}{\hbar}\, V(t) \label{i1}
\end{equation}
with $I_{\rm c}=(2 e/\hbar) E_{\rm J}$,  the phase difference across the
junction $\phi$, and where the phase velocity ($\cdot=d/dt$) is related to the
voltage drop $V(t)$ across the junction. In the opposite limit
of dominating
$E_{\rm c}$, however, charging effects prevail and Coulomb blockade is
known to rule the transfer of incoherent tunneling of Cooper pairs
\cite{nato}.

A particular appealing feature of a JJ is the fact that its classical
dynamics can be visualized as the diffusive motion of a fictitious
classical particle (RSJ model) \cite{barone}. Accordingly, for a current
biased JJ the case of very small capacitance corresponds to very strong
friction (Smoluchowski limit) associated with the phase dynamics
\begin{equation}
m \gamma \dot{\phi}+dU(\phi)/d\phi=\xi(t)\, .\label{i2}
\end{equation}
Here the translation rules are: Mass $m=(\hbar/2 e)^2 C$, damping strength
$\gamma=1/R C$, potential
 $U(\phi)=-E_{\rm J}\cos(\phi)- E_{\rm b} \phi$ with
 the energy $E_{\rm b}=(\hbar/2 e) I$ related to the bias current, and
 current noise
 $\langle \xi(t)\rangle=0$, $\langle
 \xi(t)\xi(t')\rangle=(2\gamma/\beta)\, \delta(t-t')$\,
 ($\beta=1/k_{\rm B} T$).
The corresponding time evolution equation for the phase distribution
(Smoluchowski equation) has been studied in detail already in the 60s
\cite{ivanchenk,halperin}. Even in the overdamped limit the charging
energy $E_{\rm c}$ can be relatively large and the crucial question arises
in which way charging effects alter the diffusive phase dynamics of JJs.
While this has been analyzed at very low temperatures \cite{gliprl} and
for small \cite{gli} and large \cite{kur} Josephson couplings, much less
is known for intermediate temperatures and coupling energies.

The goal of
this paper is to fill this gap for overdamped junctions by providing  a
generalization of the classical Ivanchenko-Zil'berman Theory (IZT) \cite{ivanchenk}.
As we will show, the classical and quantum regimes are
characterized by $\beta E_{\rm c}/\pi\rho\ll 1$ and $\beta E_{\rm
c}/\pi\rho\gg 1$, respectively, with $\rho=R/R_{\rm Q}\ (R_{\rm
Q}=h/4e^2)$ meaning that in a mechanical analog where
\begin{equation}
\beta E_{\rm c}/\pi\rho= \hbar\beta\gamma\, \label{i3}
\end{equation}
 the appearance of charging effects is associated with the changeover
from classical ($\gamma\hbar\beta\ll
1$) to quantum ($\gamma\hbar\beta\gg 1$) overdamped phase diffusion.
While in general the dissipative dynamics of a quantum particle is a
formidable task \cite{weiss}, it was shown recently \cite{anker1} that
 in the overdamped limit considered here the reduced quantum dynamics
 can be described by a general extension of the classical Smoluchowski
 equation, the so-called
Quantum Smoluchowski Equation (QSE). This allows to reveal the relation
between classical and Coulomb blocked transport in overdamped
JJs in terms of a simple analytical approach.

\section{Quantum phase diffusion in JJs at strong friction}

Classical overdamped motion is based upon a separation of time scales
between relaxation of position and relaxation of momentum \cite{risken}.
Momentum is slaved to position and on a coarse grained time scale can
assumed to be in thermal equilibrium with respect to the instantaneous
position of the particle.

\begin{figure}
\center \vspace*{-2.5cm}
\includegraphics[height=13cm,draft=false]{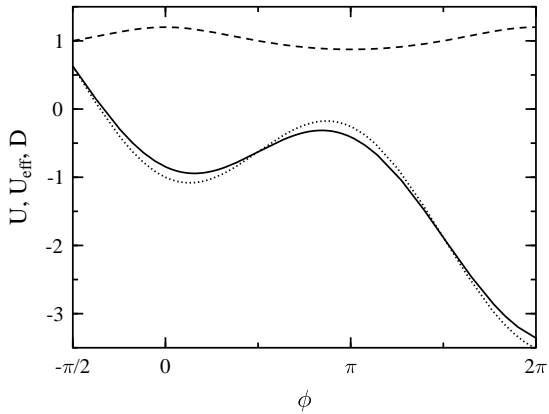}
 \vspace*{-4.7cm} \caption{Effective potential $U_{\rm
eff}(\phi)$ for $E_{\rm b}=0.4 $ (solid, in units of $E_{\rm J}$) and
diffusion coefficient $D(\phi)$ (dashed, for $\beta E_{\rm J}=1$) together
with the bare Josephson potential $U(\phi)$ (dotted, in units of $E_{\rm
J}$). Other parameters are $\beta E_{\rm c}=20$, $\rho=0.04$.}
\label{fig1}
 \vspace*{-0.0cm}
\end{figure}

Quantum dissipative dynamics is much more complicated and usually has to
be formulated in terms of the path integral formalism \cite{weiss}. This
approach has been exploited very successfully in the past to describe low
temperature properties of JJs in particular \cite{kur,gli,gliprl} and of
dissipative particles in periodic potentials in general
\cite{weiss,weiss2}. Explicit calculations require involved techniques
though, and analytic expressions are available only in certain ranges of
parameter space.  The problem is that quantum fluctuations acting on time
scales of order $\hbar\beta$ lead, particularly at low temperatures, to
long range interactions in time \cite{weiss}. Recently it was shown,
however, that in the overdamped domain the exact quantum time evolution
can be mapped onto a quantum generalization of the classical Smoluchowski
equation for the position probability distribution. The resulting QSE
reads
\begin{equation}
\frac{\partial P(\phi,t)}{\partial t}=\frac{1}{\gamma m}
\frac{\partial}{\partial \phi}\,\left[U'_{\rm
eff}(\phi)+\frac{1}{\beta}\frac{\partial}{\partial \phi} {D}(\phi)\right]\,
P(\phi,t)\,  \label{qse1}
\end{equation}
with $'=d/d\phi$ and the translation rules as specified above. Quantum
fluctuations appear as a Josephson potential with an
effective Josephson coupling
\begin{equation}
U_{\rm eff}(\phi)=-E_{\rm J}^\star\, \cos(\phi)-E_{\rm
  b}\phi\ ,\ \ \ \ \ E_{\rm J}^\star=E_{\rm J}
  \left(1-\frac{\Lambda}{2}\right)\label{qsen2}
\end{equation}
and a phase dependent diffusion coefficient capturing quantum noise\cite{remark}
\begin{equation}
 D(\phi)=[1-\theta\cos(\phi)]^{-1}\ ,\ \ \ \ \ \theta=\Lambda\beta
 E_{\rm J}\, .
\label{qsen3}
\end{equation}
The changeover from classical to quantum dynamics is governed by a
function $\Lambda$ that contains $\gamma\hbar\beta$ as the essential
quantity. With  (\ref{i3}) it takes the explicit form
\begin{equation}
\Lambda=2 \rho\, \left[c+\frac{2 \pi^2 \rho}{\beta E_{\rm
c}}+\Psi\left(\frac{\beta E_{\rm c}}{2 \pi^2\rho }\right)\right]\,
\label{qsen3b}
\end{equation}
where $c=0.5772\ldots$ is
  Euler's constant and  $\Psi(\cdot)$ denotes the logarithmic derivative of
the gamma function. In the high temperature limit $\beta E_{\rm c}/\rho\ll
1$ quantum fluctuations become small and independent of $\rho$, i.e.,
  $\Lambda\approx\beta E_{\rm c}/\pi^2$,
and one regains classical Smoluchowski dynamics. In the low temperature
range $\beta E_{\rm c}/\rho\gg 1$ one has $\Lambda\approx
2\rho\log(\beta E_{\rm c}/2 \pi^2\rho )$ leading to a substantial
  influence of quantum fluctuations.
Qualitatively, the effects of  $E_{\rm J}^\star$ and $\theta$
on the dynamics of the fictitious particle are easily understood
(cf.~fig.~\ref{fig1}): $E_{\rm J}^\star<E_{\rm J}$ lowers the potential
barrier in $U_{\rm eff}$, $\theta$-dependent terms in the  diffusion
coefficient [see
  (\ref{qsen2},\ref{qsen3})]
enhance the noise strength in the wells ($D>1$) and lower it around the
barrier tops ($D<1$). It turns out that this causes a moving particle
to  experience a larger mobility where in limiting cases barrier
  ($\beta E_{\rm J}\ll 1$) or diffusion  ($\beta E_{\rm J}\gg 1$)
  related quantum fluctuations dominate.

Let us now first address the range of applicability of (\ref{qse1}) for
the physics of JJs. The QSE is valid if (i) a separation of time scales
$\hbar\beta, 1/\gamma\ll \gamma/\omega_{\rm J}^2$ with the plasma
frequency $\omega_{\rm J}=(2 e/\hbar) R I_{\rm c}$ is guaranteed.
Classically, for the McCumber parameter this condition reduces to $\beta_{\rm
  J}=\gamma/\omega_{\rm J}\gg 1$.
Further, (ii) in order for the momentum $m\dot{\phi}$ to relax within the
$RC=1/\gamma$-time to a Boltzmann distribution around $\langle
\dot{\phi}\rangle$, the external voltage $V$ is restricted by $e V\ll
\hbar\gamma$. We note in passing that (ii) also ensures that the actual
non-ohmic impedance seen by the junction can effectively be treated as
ohmic. By combining (i) and (ii) and re-expressing them in junction
parameters
 we thus expect the QSE to describe quantum phase diffusion in JJ if
 (see fig.~\ref{fig2})
\begin{equation}
\frac{E_{\rm c}}{E_{\rm J}2\pi^2\rho^2}\gg
  1,\ \pi\rho\beta E_{\rm J},\ \frac{V}{R I_{\rm c}}\,
  .\label{qsen4}
\end{equation}
Since typically $\rho\ll 1$ the above condition allows for a broad range
of values for $E_{\rm c}/E_{\rm J}$,  $\beta E_{\rm J}$, and also large
voltages $V/R I_{\rm c}$. It is important to note that for realistic JJ
with finite $\rho$ the classical description \cite{ivanchenk,halperin} is
always restricted to small and moderate values of $\beta E_{\rm J}$,
while with increasing $\beta E_{\rm J}$ one approaches the quantum
Smoluchowski range with a finite $\Lambda$ [see
(\ref{qsen3b}) and fig.~\ref{fig2}].

To complete this discussion we also specify the corresponding quantum
Langevin equation (in the Ito sense) which generalizes the classical one
(\ref{i2}) to
\begin{equation}
m \gamma \dot{\phi} + U'_{\rm eff}(\phi)=\xi(t)\ \sqrt{D(\phi)}\, .\label{qse6}
\end{equation}
Now, the Eqs.~(\ref{qse1}) and (\ref{qse6}) are the starting point to
study in a very elegant way all aspects of the phase dynamics of JJs
within the range (\ref{qsen4}). While in the quantum domain very low
temperatures ($T\to 0$) have been explored in detail \cite{gliprl}, finite
temperature results exists only for small \cite{gli} and large
\cite{kur} Josephson
couplings, respectively. In the sequel we will show that the
QSE bridges between these limiting cases.

\section{Steady state current and current--voltage characteristics}

For finite bias current $I_b>0$ the phase dynamics (\ref{qse1})
approaches a stationary non-equilibrium state at longer times. This is
associated  with a stationary current $J_{\rm st}=\lim_{t\to
\infty}\langle \dot{\phi}(t)\rangle/2\pi$ given by
\begin{equation}
J_{\rm st}=\frac{1-{\rm e}^{-\beta 2\pi E_{\rm b}}}{\gamma
m\beta}\left[\int_0^{2\pi}d\phi\,  \frac{{\rm
e}^{-\beta{\psi}(\phi)}}{{D}(\phi)} \, \int_\phi^{\phi+2\pi} d\phi'\, {\rm
e}^{\beta{\psi}(\phi')}\right]^{-1}\label{qse3}
\end{equation}
with
\begin{equation}
\psi(\phi)=-E_{\rm J}^\star \cos(\phi)-E_{\rm b} \phi+\theta E_{\rm b}
\sin(\phi)-\frac{1}{2} \theta\, E_{\rm J}^\star\sin^2(\phi)\,
.\label{curr4}
\end{equation}
According to  $\langle V\rangle=(h/2e) J_{\rm st}$ one then finds the
current voltage--characteristics of a current biased junction to read
\begin{equation}
\langle V\rangle = \frac{ \rho\, \pi}{\beta e}\frac{1-{\rm e}^{-2\pi
\beta E_{\rm b}}}{T_{\rm qm}}\, \label{curr5}
\end{equation}
where the nominator $T_{\rm qm}$ results from normalizing the steady state
phase distribution to 1 and can be written as
\begin{equation}
T_{\rm qm}=\frac{1}{2\pi}\int_0^{2\pi}d\phi \int_0^{2\pi} d\phi'\, {\rm
e}^{-\beta E_{\rm b} \phi}\,  {\rm e}^{-2\beta E_{\rm
J}^\star\cos(\phi')\sin(\phi/2)}\, [1-\theta \sin(\phi'-\phi/2)]\,  {\rm
e}^{2\beta\theta \xi(\phi,\phi')}\label{curr6}
\end{equation}
with
\begin{equation}
\xi(\phi,\phi')=  \sin(\phi')\sin(\phi/2)\left[E_{\rm b}+ E_{\rm J}^\star
\cos(\phi') \cos(\phi/2)\right]\, .\label{curr7}
\end{equation}
The expression (\ref{curr5}) together with (\ref{curr6}) is the central
result and will be discussed in the following.
\begin{figure}
\center \vspace*{-2.5cm}
\includegraphics[height=13cm,draft=false]{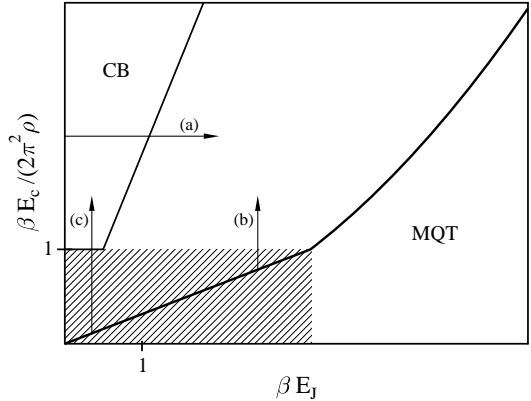}
 \vspace*{-5cm}
 \caption{Range of the QSE  for a
 JJ with $\rho\ll 1$,  $V/R I_{\rm c}<1$. The classical range
 (shaded)
 and the domains of Coulomb blockade (CB) and
 macroscopic quantum tunneling (MQT) are indicated. The QSE is applicable above the
 thick line, see (\ref{qsen4}), and the arrows illustrate various
 changeovers discussed in the text. The slope of the CB boundary decreases
 with increasing $V/R I_{\rm c}$ so that for $V/R I_{\rm c}>1$ CB dominates in
 most parts of parameter space.}\label{fig2}
 \vspace*{-0.0cm}
\end{figure}

\begin{figure}
\center \vspace*{-5.5cm}
\includegraphics[height=16cm,draft=false]{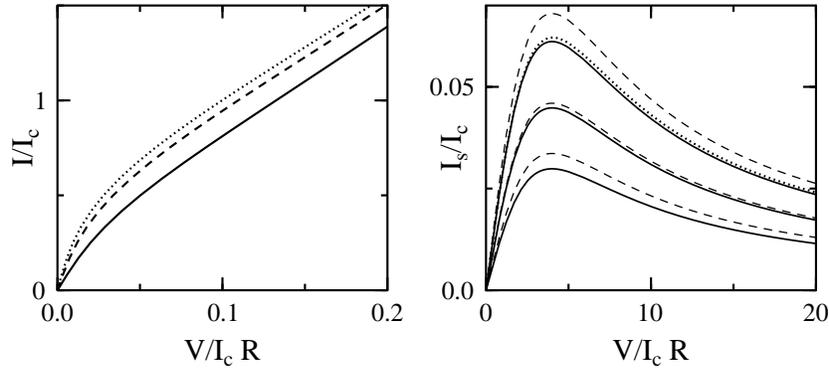}
 \vspace*{-5.2cm}
  \caption{Left: Current-voltage characteristics for
 $\beta E_{\rm J}=2$, $\rho=0.04$. The quantum results [$\beta
E_{\rm c}=1$ (dashed), $\beta E_{\rm c}=20$ (solid)] are shown with the
classical one ($\beta E_{\rm c}=0$, dotted). Right: Supercurrent vs.
voltage for $\beta E_{\rm J}=0.25$, $\rho=0.04$.  The classical result
(dotted, \cite{ivanchenk}) and the CB expression [dashed, (\ref{super5})]
are depicted together with the full QSE result (solid) for
$\beta E_{\rm c}=0.15$ (top), $\beta E_{\rm c}=20$ (middle), $\beta E_{\rm
c}=1000$ (bottom).} \label{fig3}
 \vspace*{-0.0cm}
\end{figure}

For $\beta E_{\rm c}/\rho\ll 1$ the function
$T_{\rm qm}$ reduces to the classical result
\begin{equation}
T_{\rm cl}(E_{\rm J})=\int_0^{2\pi} d\phi'\, {\rm e}^{-\beta E_{\rm b}
\phi'}\, I_0[2\beta E_{\rm J}\sin(\phi'/2)]\label{curr8}
\end{equation}
with $I_0$ the modified Bessel function \cite{barone} and IZT
\cite{ivanchenk,halperin} is recovered. On the other hand, in the low
temperature domain $\beta E_{\rm c}/\rho\gg 1$ two distinct ranges must be
distinguished (see fig.~\ref{fig2}). For smaller couplings, $\beta E_{\rm
  J}<1$, where Coulomb blockade (CB) transfers charges incoherently,
we have $\theta\ll 1$ so that diffusion related quantum
fluctuations are negligible. Then, (\ref{curr6}) takes the classical form
with quantum fluctuations incorporated in terms of a renormalized
Josephson energy, i.e.\ $T_{\rm
  qm}\approx T_{\rm cl}(E_{\rm J}^\star)$. This important extension of
IZT has first been derived in
\cite{gli} based on  a direct evaluation of the real-time path integral
expression. However, since diffusion related quantum fluctuations are
disregarded, this simple result is (for finite $\rho\ll 1$) indeed restricted to
 $\beta E_{\rm J}{\textstyle {\lower 2pt \hbox{$<$} \atop \raise 1pt
    \hbox{$ \sim$}}}1$. In
contrast, (\ref{curr6}) also holds for much larger $E_{\rm J}$
[fig.~\ref{fig2}, arrow (a)] associated with coherent Cooper pair
tunneling. An explicit formula can be obtained from (\ref{curr5}) for
sufficiently large $\beta E_{\rm J}>1$ and $\alpha=E_{\rm b}/E_{\rm
J}=I/I_{\rm c}<1$. Then, occasional phase slips lead to the voltage
\begin{equation}
\frac{\langle V\rangle}{R I_{\rm c}}=\frac{\sqrt{1-\alpha^2}}{2\pi}\ {\rm
    e}^{ -2\beta E_{\rm J}
    (1-\alpha^2)^{3/2}/(3\alpha^2)}\ {\rm e}^{ 2\theta
    \sqrt{1-\alpha^2}}\, \label{curr8b}
\end{equation}
which via $\theta$ is affected by diffusion related quantum
fluctuations. Accordingly,  quantum effects {\em cannot} always be captured
by a renormalization of the Josephson coupling, rather such a
simplification occurs for CB dominated
transport only. As can be observed in fig.~\ref{fig2}, the result
(\ref{curr8b}) tends for $\theta\to 0$ to classical thermal
activation over the barriers of the washboard potential $U(\phi)$
where quantum corrections are of the known damping independent form \cite{feynman}.
At lower temperatures, i.e.\ for finite $\theta$, they show a
complicated dependence on $\rho$ and capture the precursors of macroscopic quantum
tunneling (MQT) found at very low temperatures \cite{kur}. Thus, the
central result  (\ref{curr5}) fills the gap between established results in
different transport domains: On the one hand for fixed $\beta E_{\rm
c}/\rho>1$ it leads with increasing $\beta E_{\rm J}$ from Coulomb
blockade to coherent Cooper pair tunneling [fig.~\ref{fig2}, arrow (a)].
On the other hand
 for fixed $\beta E_{\rm J}>1$ it connects with varying $\beta E_{\rm c}/\rho$
 classical thermal activation with MQT [fig.~\ref{fig2}, arrow (b)]. Apart
 from limiting cases, the result (\ref{curr5}) must be evaluated numerically as
 illustrated in fig.~\ref{fig3} (left).

What is left, is the domain of small Josephson coupling $\beta E_{\rm
J}<1$ and increasing $\beta E_{\rm c}/\rho$ [fig.~\ref{fig2}, arrow (c)]
that has been discussed in the literature \cite{gli,gliprl} for a voltage
biased junction. This changeover leads from classical coherent IZT
\cite{ivanchenk} to CB \cite{udo}. Here, we demonstrate that the QSE recovers this
scenario, too, and hence gives a complete description throughout the QSE
range sketched in fig.~\ref{fig2}. Results for a voltage biased JJ are
gained from the
current biased case  by means of $I=V/R$ and $\langle V\rangle=V-R I_{\rm
s}$. The interesting observable then is the supercurrent $I_{\rm s}$. From
(\ref{curr5}) we obtain to order $(\beta E_{\rm J})^2$
\begin{equation}
\frac{I_{\rm s}}{I_{\rm c}}=\frac{(\beta E_{\rm J}^\star)^2}{2}\,
\frac{V}{R I_{\rm c}}\ \frac{1}{1+[\beta E_{\rm J} V/(R I_{\rm c})]^2}\,
\label{super3}
\end{equation}
with the characteristic renormalized $E_{\rm J}$ discussed above. This
result coincides with that of \cite{gli} and connects
quite different transport regimes [see figs.~\ref{fig2},\ref{fig3}
(right)]. Namely, in the classical limit $\beta E_{\rm c}/\rho\ll 1$  it
gives the small-$\beta E_{\rm J}$-limit of the IZT \cite{ivanchenk},
while in the opposite quantum domain  $\beta
E_{\rm c}/\rho\gg 1$ it agrees with the small $\rho, \Lambda$ expansion
of the CB expression \cite{udo}
\begin{equation}
\frac{I_{\rm s}}{I_{\rm c}}=\frac{\beta E_{\rm J}\, {\rm
e}^{-\Lambda}}{4\pi}\ \frac{|\Gamma(\rho-i\beta e
V/\pi)|^2}{\Gamma(2\rho)}\, {\rm sinh}(\beta e V)\, .\label{super5}
\end{equation}
Thus, we conclude that the quantum Smoluchowski result (\ref{super3})
interpolates in the domain of small Josephson couplings between
phase-coherent and incoherent charge transport (with $E_{\rm J}\to E_{\rm
J}^\star$) where the strength of charging effects controls the changeover
from one regime to the other (see fig.~\ref{fig2}). Note that
(\ref{super3}) yields the supercurrent reasonably well also for $\beta
E_{\rm c}\gg 1$ ($\Lambda$ of order 1) [fig.~\ref{fig3} (right), bottom].

\section{Conclusions}
Based on the analogy between JJ physics and the diffusive motion of a
fictitious particle in a washboard-type of potential we generalized for
strong friction (low capacitance) the IZT  to the low temperature
quantum domain. Quantum fluctuations affect
the potential profile as well as the diffusion constant and physically
are associated with charging phenomena. This formulation reduces to known
results derived in certain limits and thus gives a complete description
over broad domains of parameter space. Our main expression is specified
in (\ref{curr5}) together with (\ref{curr6}). It contains incoherent
charge transfer (CB) and coherent Cooper pair tunneling as well as the
precursors of MQT. In particular, the incorporation of quantum
fluctuations in terms of a renormalized Josephson junction turned out to
be characteristic for CB only. Eventually, our findings confirm the
recently developed quantum Smoluchowski theory to be a powerful tool for
studying low temperature quantum diffusion in mesoscopic physics.

\acknowledgments We thank H. Grabert for motivating discussions and
gratefully acknowledge financial support of the DFG (Bonn) through SFB276.

\end{document}